\documentclass[sigconf]{acmart}
\usepackage{graphicx}
\usepackage{booktabs} 
\usepackage{csvsimple}
\usepackage{multirow}
\usepackage{courier}

\setcopyright{rightsretained}

\acmDOI{xxxx}

\acmISBN{xxxx}

\acmConference[CASCON'18]{28th Annual International Conference on Computer Science and Software Engineering}{October 2018}{Canada}
\acmYear{2018}
\copyrightyear{2018}

\acmArticle{4}
\acmPrice{15.00}

\begin{document}
\title{Empirical Vulnerability Analysis of Automated Smart Contracts Security Testing on Blockchains}

\author{Reza M. Parizi}
\orcid{1234-5678-9012}
\affiliation{%
  \institution{Department of Software Engineering and Game Development, Kennesaw State University}
   \city{GA}
  \state{USA}
}
\email{rparizi1@kennesaw.edu}

\author{Ali Dehghantanha}

\affiliation{%
  \institution{School of Computer Science, University of Guelph}
   \city{Ontario}
  \state{Canada}
}
\email{a.dehghan@uoguelph.ca}

\author{Kim-Kwang Raymond Choo}

\affiliation{%
	\institution{Department of Information Systems and Cyber Security, University of Texas at San Antonio}
	\city{Texas}
	\state{USA}
}
\email{raymond.choo@fulbrightmail.org}

\author{Amritraj Singh}
\affiliation{%
	\institution{Department of Software Engineering and Game Development, Kennesaw State University}
	\city{GA}
	\state{USA}
}
\email{amritra@students.kennesaw.edu}

\renewcommand{\shortauthors}{Reza M. Parizi et al.}

\begin{abstract}
The emerging blockchain technology supports decentralized computing paradigm shift and is a rapidly approaching phenomenon. While blockchain is thought primarily as the basis of Bitcoin, its application has grown far beyond cryptocurrencies due to the introduction of smart contracts. Smart contracts are self-enforcing pieces of software, which reside and run over a hosting blockchain. Using blockchain-based smart contracts for secure and transparent management to govern interactions (authentication, connection, and transaction) in Internet-enabled environments, mostly IoT, is a niche area of research and practice. However, writing trustworthy and safe smart contracts can be tremendously challenging because of the complicated semantics of underlying domain-specific languages and its testability. There have been high-profile incidents that indicate blockchain smart contracts could contain various code-security vulnerabilities, instigating financial harms. When it involves security of smart contracts, developers embracing the ability to write the contracts should be capable of testing their code, for diagnosing security vulnerabilities, before deploying them to the immutable environments on blockchains. However, there are only a handful of security testing tools for smart contracts. This implies that the existing research on automatic smart contracts security testing is not adequate and remains in a very stage of infancy. With a specific goal to more readily realize the application of blockchain smart contracts in security and privacy, we should first understand their vulnerabilities before widespread implementation. Accordingly, the goal of this paper is to carry out a far-reaching experimental assessment of current static smart contracts security testing tools, for the most widely used blockchain, the Ethereum and its domain-specific programming language, Solidity, to provide the first body of knowledge for creating more secure blockchain-based software. 
\end{abstract}



\maketitle
\section{Introduction}
Smart Contracts (SC) have grown in popularity in the recent past years and are believed to be the next generation of automation in inter-party agreements in the blockchain-based systems. Smart contracts are self-directing agreements implemented through a piece of software whose autonomous execution applies the terms of the settlement and measurements. The main idea behind SC is to urge obviate traditional trusted third parties (authority, entity or organization) to be replaced by pieces of code running on a decentralized and immutable system. This new paradigm of SC applications opens the door to many opportunities. One of the promising areas is for the IoT security and forensics which has drawn a lot of attention from both academia and the enterprise world \cite{wang2018a, jhaveri2008a, gao2018privacy, zhang2018a, li2018a, epiphaniou2017a}. In fact, the implementation and use of blockchain has far surpassed its original intended purpose as the backbone to the world's first decentralized cryptocurrency. The value of a trustless, decentralized ledger that carries historic immutability has been recognized by other industries looking to apply the core concepts to existing business processes. These unique properties of the blockchain make its application an attractive idea for many areas of business, including IoT. The major issue in IoT security is to control and know `who' will be connecting to the network across a large number of things (e.g. sensors and devices) without breaching data privacy \cite{conti2018a}.  Decentralized smart contracts on blockchains seem to be a vital remedy \cite{zhang2018b}, \cite{christidis2016a}, \cite{andersen2017a}, especially in dealing with security flaws in widely distributed IoT nodes \cite{azmoodeh2018a}. The blockchain technology \cite{yong2016a} is the essential means for delivering this trust model envisioned by smart contracts, and appears to hold great promise for future IoT advancement.

Although blockchain technology is perceived as secure by design, its embedded applications (i.e. SC) in dynamic environments (such as IoT) may introduce vulnerabilities in real life situations \cite{pajouh2016a}. After all, these smart contract applications controlling nodes and transactions are snippets of code written by fallible human developers. Besides, because of their special nature, mistakes or bugs can have a significant financial impact, hence security is of utmost importance.

To date, smart contracts have been affected by unfortunate incidents and attacks (e.g., reentrancy problem in 'splitDAO' function caused \$40 million loss in June 2016 \cite{atzei-a}, and \$32 million was stolen by attackers due a bug within the code in November 2017 \cite{destefanis2018a}). These high-profile incidents demonstrate developers (even those with experiences) may put behind security bugs in smart contracts that could create serious vulnerabilities for attackers' misuse \cite{parizi2018b}. This would be further extended in IoT environments because of their sheer size and velocity \cite{watson2016a}. Thus, writing trustworthy and safe smart contracts can be extremely difficult due to the complicated semantics of underlying domain-specific languages and its testability. One effective way to mitigate this issue is, every developer embracing the ability to write their smart contracts should be capable of testing their own code, against vulnerabilities, via automated static code analysis tools \cite{chess2004a}, \cite{parizi2018a}, and before deploying them to the immutable environments on blockchains. Being aware of the security pitfalls involved with writing smart contracts as a preventive strategy to avoid known mistakes and vulnerabilities is a smart choice, since the cost of failure on a blockchain that handles an IoT network can be very high. Currently, the state of empirical knowledge within the area of smart contracts security is in embryonic form. Thus, the aim of this paper is to take this initiative by providing a first-time empirical evaluation of static smart contracts security testing tools, for Ethereum as the most widely used blockchain in the community and its popular domain-specific programming language, Solidity. As the software and security engineering research community has long relied on Free and Open Source Software (FOSS) tools for security testing, our main emphasis would be on open source tools as well. With this empirical analysis, the objective of the research is to assess the four FOSS tools (namely Oyente \cite{luu-a}, Mythril \cite{Mythril}, Securify \cite{secu}, and SmartCheck \cite{sdek}) based on their vulnerability detection effectiveness and the accuracy. To accomplish the stated objective, we ponder to examine the accompanying research questions:

\begin{itemize}
\item \textbf{RQ1} - How effective are the automated smart contract security testing tools, in terms of vulnerability detection ability? And what is the most effective tool?

\item \textbf{RQ2} - What are the accuracy scores obtained by the automated testing security tools in detecting true vulnerabilities?
\end{itemize}

To manage RQ1 and RQ2, we carried out large-scale experiments among the aforementioned tools. Based on the presented research questions, the subsequent hypotheses were detailed: 

\begin{itemize}
\item \textit{Null hypothesis ($H_{0Eff}$)} = There is no significant difference in vulnerability detection effectiveness between the four smart contract security testing tools. It can be expressed as:
$H_{0Eff} : \mu_{Oyente} = \mu_{Mythril} = \mu_{Securify} = \mu_{SmartCheck}$. Where, $\mu_t$ is the mean detection effectiveness of testing tool \textit{t} measured on all the participating smart contracts (as shown in Eq. (\ref{one})).

\item \textit{Alternative hypothesis ($H_{1Eff}$)} = There is significant difference in vulnerability detection effectiveness between the four automated tools. It can be expressed as: Given \textit{z} = \{Oyente, Mythril, Securify, SmartCheck\}, \textit{x} and \textit{y} as two single-valued variables belonging to \textit{z}, therefore, $H_{1Eff} : \exists_x|\mu_x > \mu_{y\epsilon z-x}$.

\item \textit{Null hypothesis ($H_{0Acc}$)} = There is no significant difference in accuracy scores between the four automated tools. It can be expressed as:
$H_{0Acc}:  \theta_{Oyente} = \theta_{Mythril} = \theta_{Securify} = \theta_{SmartCheck}$. Where, $\theta_t$ is the mean of the accuracy of testing tool \textit{t} measured on all the participating smart contracts (as shown in Eq. (\ref{two})).

\item \textit{Alternative hypothesis ($H_{1Acc}$)} = There is significant difference in accuracy scores between the four automated tools. It can be expressed as: Given \textit{z} = \{Oyente, Mythril, Securify, SmartCheck\}, \textit{x} and \textit{y} as two single-valued variables belonging to \textit{z}, therefore, $H_{1Acc} : \exists_x|\theta_x > \theta_{y\epsilon z-x}$.
\end{itemize}

Within our exploration, there is only one independent variable, i.e. the security testing tool designed for smart contracts. For this main factor there are four treatments (tools): \textit{Oyente, Mythril, Securify, SmartCheck} (See Section 2.1). The vulnerability detection effectiveness and the accuracy are the dependent variables measured in this work. We are collecting the following four building-block metrics to measure the dependent variables: 

\begin{itemize}
\item True Positive - $\mathbf{TP_c}$: tool correctly identifies a real vulnerability in contract \textit{c}.

\item False Negative - $\mathbf{FN_c}$: tool fails to identify a real vulnerability in contract \textit{c}.

\item True Negative - $\mathbf{TN_c}$: tool correctly ignores a false alarm in contract \textit{c}.

\item False Positive - $\mathbf{FP_c}$: tool fails to ignore a false alarm in contract \textit{c}.
\end{itemize}	 

\begin{table*}[]
  \centering
  \caption{Summary of current tools and their characteristics}
    \begin{tabular}{p{5em}p{5em}p{5em}p{5em}p{25em}}
    \toprule
    \textbf{Tool } & \textbf{Level of rigor} & \textbf{Analysis basis} & \textbf{Interface Means} & \textbf{Description} \\
    \midrule
    Oyente & Heuristic & source code (.sol) & Command line-based tool & Oyente \cite{luu-a} is an automated security analysis tool for revealing security vulnerabilities in smart contracts. \\
    \hline
    Mythril & Analytic and Heuristic & Byte code and source code (.sol) & Command line-based tool & Mythril \cite{Mythril} is an automated security analysis tool for Ethereum smart contracts. It uses Concolic analysis, taint analysis and control flow checking to detect a variety of security vulnerabilities. The analysis is based on laser-Ethereum, a symbolic execution library for EVM bytecode. \\
    \hline
    Securify & Formal & Byte code and source code (.sol) & UI-based tool & Securify \cite{secu} is an automated formal security analysis tool for Ethereum smart contracts. It can detect various security issues such as input validation, reentrancy, and others. \\
    \hline
    SmartCheck & Analytic and Heuristic & source code (.sol) & UI-based tool & SmartCheck  \cite{sdek} is an automated static code analyzer developed by SmartDec Security Team. It runs analysis in Solidity source code and automatically checks smart contracts for security vulnerabilities and bad practices. \\
    \bottomrule
    \end{tabular}%
  \label{tab:addlabel}%
\end{table*}%

The vulnerability detection effectiveness of a tool is measured as following (which is also referred to as Recall \cite{salton1988a} or sensitivity): 

\begin{equation}
Eff_j = (\sum_{i=1}^n \frac{\# TP_i}{\# TP_i + \# FN_i}/n) * 100 \label{one}
\end{equation}

Where \textit{n} is the number of smart contracts used, and \textit{j} represents a given tool (i.e., \textit{j} = 1, 2, 3, 4). The accuracy score of a tool is measured as following (which is essentially a Youden Index \cite{youden1950a}):

\begin{equation}
Acc_j = (Eff_j + ((\sum_{i=1}^n \frac{\# TN_i}{\# TN_i + \# FP_i}/n) * 100)) - 1 \label{two}
\end{equation}

The rest of this paper is organized as follows. Section 2 provides an overview of the current tools and background in smart contracts. Section 3 gives and describes the experimental setup; Section 4 gives experimental design and validity. Section 5 analyzes the obtained data and discusses the results; Section 6 presents the related work in the literature; and finally, Section 7 reports the conclusions.

\section{Background}
The concept of smart contract was initially perceived at the source code level, `code is law'.  The code-based testing not only in blockchain programming (including smart contracts development) but also in other computing paradigms is still favored and effective as it determines a reasonable degree of reliance on the most comprehensive artifact of development process, and before deploying it to the hosting environment \cite{parizi2015a, parizi2009a}. In addition, due to the special nature of smart contracts on the blockchain, static security analysis prior to deployment seems to be the perfect fit. Inspired by this, the paper primarily focuses on the automated static smart contracts security tools (see Section 2.1) in the Ethereum blockchain by focusing its attention on Solidity (see Section 2.2).

\subsection{Automated Smart Contract Security Testing Tools}
There are a handful of tools for automated smart contract (written in Solidity) security vulnerability testing based on code-level analysis. We give a synopsis of the four most related FOSS tools that we used in our experiments, namely Oyente \cite{luu-a}, Mythril \cite{Mythril}, Securify \cite{secu}, and SmartCheck \cite{sdek}. Table 1 shows the summary of the selected tools and their main properties. Level of rigor, ranging from syntactic, heuristic, analytic to fully formal, refers to underlying security testing technique of the given tool. The Analysis basis column represents the type of artifact (source code, byte code, or binary) that a tool uses as its input source.

\subsection{Solidity, EVM, and Ethereum Smart Contracts}
Ethereum is a consensus-based framework that utilizes the blockchain technology to offer a globally open decentralized computing platform, referred to as Ethereum Virtual Machine (EVM). EVM programs are implemented in bytecode that operates on a simple stack machine supporting a handful of instructions. Contracts reside on the blockchain in an Ethereum-specific binary format (EVM bytecode). Developers, however, do not usually write low-level EVM codes for contracts. Instead, they typically use a high-level language in an exceedingly JavaScript-like syntax called \textit{Solidity} that compiles into bytecode to be uploaded on the blockchain. Ethereum is gaining a substantial popularity in the blockchain community, since the major purpose of EVM programs is as smart contracts that manage digital assets \cite{bhargavan2016a}, and for creating decentralized applications (dApps).

Solidity \cite{eth} is a domain and platform-specific language that is designed for writing smart contracts on the Ethereum blockchain. It has been the most widely used open-source programming language in implementing public and private blockchains \cite{ethereum}. The syntax of Solidity is analogous to ECMAScript (JavaScript), where the resulting code associated with a smart contract is compiled to bytecode and then is run on the EVM to execute the functionality of the contract.   

Similar to different blockchain platforms (e.g., Bitcoin, Hyperledger Fabric), Ethereum provides a peer-to-peer (P2P) network for the users. The underlying Ethereum blockchain database is supported and updated by the participating nodes connected to the current network. Each node on the network runs the EVM and executes identical set of directions. The Ethereum platform itself is featureless or value-agnostic. It is up to organizations and developers to choose its utilization in view of their business objectives. In any case, certain application composes assortments benefit more than others from Ethereum's capabilities. In particular, Ethereum is fitted to applications that automate coordinate association between peers or encourage facilitated bunch activity over a network. For example, applications for planning digital commercial centers, or the computerization of complex financial related contracts. With regards to programming on Ethereum, there are some key focuses to note from the Ethereum Design Rationale (EDR) document \cite{desrat}. 

In the context of Ethereum, a typical smart contract is recognized by a unique \textit{contract address} rendered after a successful production of a creation transaction. A blockchain state is hence a mapping from addresses to nodes' accounts in the network, where each smart contract account holds an associate quantity of \textit{virtual coins} (Ether - the contract balance) by possessing its own non-public \textit{state} and \textit{storage}. 

Figure 1 shows an illustrative sample for a user-defined smart contract, named \texttt{Puzzle} \cite{luu-a}. This contract manages a simple reward system in which whoever solves a problem (i.e. puzzle) will get rewards. To make this contract up and running, the EVM bytecode of the contract is first sent to miners through a \textit{contract-creation} transaction. Once the transaction is admitted into a block (has to pass through mining process) in the hosting blockchain, a unique address for Puzzle is generated. Then each miner instantiates the contract by executing its constructor (Line 8), and a local storage is assigned in the blockchain. Finally, the main code implemented in the anonymous function of Puzzle (Lines 15) is added to the contract's storage.

\begin{figure}[h]
\centering
\includegraphics[scale=1]{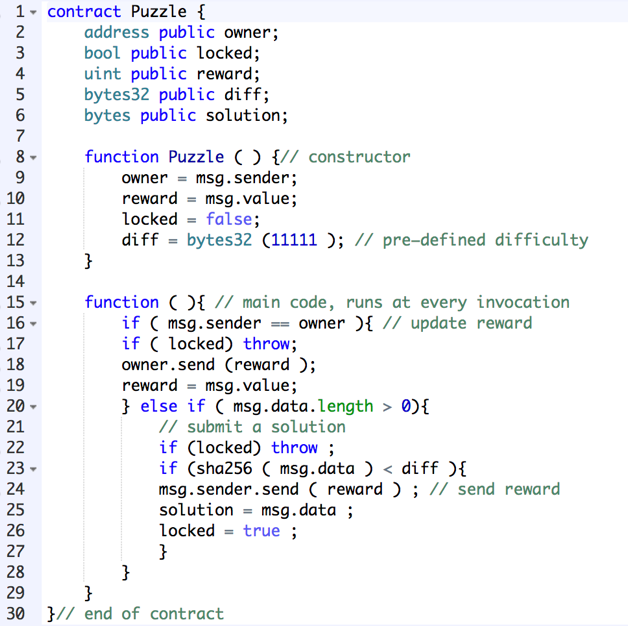}
\caption{Smart contract example in Solidity}
\end{figure}

Once the contract is live on the blockchain, it can be used to deliver its service to users. Whenever a user submits his/her solution, a \textit{contract-invoking} transaction gets directed to the associated address of the contract. This in turn triggers the execution of the main function defined at Line 15. In this case, all the sender's information, the Ether amount, and the input data of the invoking transaction will be stored in the \textit{msg} variable. The contract then processes the receiving data and handles the reward according to its business logic written in the main function. When a correct solution is received, the contract automatically redeems the reward to the sender by executing the code at Line 24.

\section{Experimental Setup}
The general process of experiments is depicted in Figure 2. The operation process starts with choosing a security testing tool in a random manner, i.e. treatment, following by acquiring a chosen random smart contract from the pool of contracts. 

\begin{figure}[h]
\centering
\includegraphics[scale=0.3]{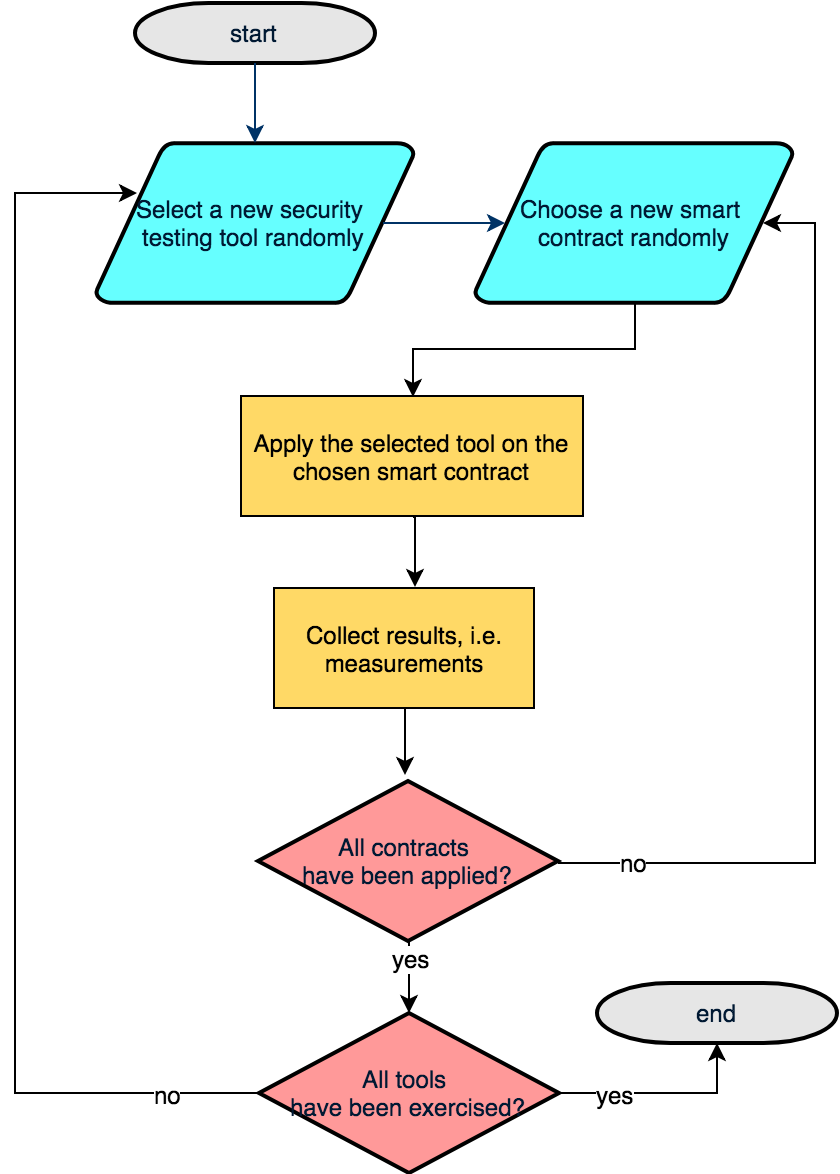}
\caption{The general experiment execution process}
\end{figure}

Tools without a Web-based UI were run from their source files on an online compiler, i.e. Remix\footnote{http://remix.ethereum.org/\#optimize=false\&version=soljson-v0.4.21+commit.dfe3193c.js.}  as Solidity IDE, or \textit{solc} command-line Solidity compiler - whichever possible. We followed the instructions on tools' respective GitHub pages to configure installation. Prior to the experiment, we compiled all contracts and made sure they were all free from compilation errors. 

After running the tools against each smart contract, we recorded the security threats they raised, and assessed whether or not they accurately identified the known vulnerabilities, given the four-fundamental metrics of classification, TP, FN, TN, FP. 

As with any experimental setting, we were required to collate a set of object programs (in our case `vulnerable smart contracts') through observation of experiments execution process. One of the challenges we faced was the difficulty of finding representative contracts to use in our study without getting involved in legal and privacy issues. Historically, authors or developers have seen the availability of deliberately vulnerable products (such as vulnerable Website for Web Apps security testing) as a practical solution to pursue their studies without the fear of prosecution. In this spirit and after an exhaustive search in the online resources, we found two suitable publicly-available sources of intentionally vulnerable contracts: (1) contracts developed by Trail of Bits\footnote{https://www.trailofbits.com/} , a dedicated security team on foundational tools and deep expertise in cryptography and Ethereum security; (2) contracts developed by the Ethernaut\footnote{https://ethernaut.zeppelin.solutions/}   wargame, an online smart contract hacking challenge by Zeppelin. Although one of the participating tool (Mythril) comes with a collection of test contracts\footnote{https://github.com/ConsenSys/mythril/tree/master/solidity\_examples}, we decided not to use those with a specific end goal to make the evaluation fair-minded and to give a reasonable equivalent platform for analyzing all the tools.

Afterwards, we selected a suit of ten contracts written in Solidity, as presented in Table 2. These contracts encompass a wide range of real-world vulnerabilities (including Integer Overflow, Missing Constructor, Reentrancy, Unchecked External Call, Unprotected Function, Wrong Interface, Callstack Depth Attack, Assertion Failure, Timestamp Dependency, Parity Multisig Bug, Transaction-Ordering Dependence (TOD, etc.) with many instances for each class of vulnerability. In this way, the outcomes started from these contracts for the proposed experiments will give premise to drawing more convincing bits of knowledge. We intentionally only chose \textit{open source} contracts to ensure that our experiments can be replicated by public access to the source code.

\begin{table}[htbp]
  \centering
  \caption{Smart contracts used in the experiment.}
    \begin{tabular}{cccc}
    \toprule
    \textbf{Smart contracts} & \textbf{\# function} & \textbf{\#LOC} & \textbf{Source} \\
    \midrule
    Fallback & 5     & 33    & Ethernaut \\
    Fallout & 5     & 31    & Ethernaut \\
    Token & 3     & 20    & Ethernaut \\
    King  & 2     & 21    & Ethernaut \\
    Re-entrancy & 4     & 25    & Ethernaut \\
    RaceCondition & 9    & 50    & Trail of Bits \\
    Rubixi & 18    & 155   & Trail of Bits \\
    GiftBox & 7     & 69    & Trail of Bits \\
    KingOfTheEtherThrone & 7     & 170   & Trail of Bits \\
    WalletLibrary & 44    & 463   & Trail of Bits \\
    \bottomrule
    \end{tabular}%
  \label{tab:addlabel}%
\end{table}%

All experiments were conducted on a similar computer (to maintain a strategic distance from ecological predisposition) utilizing \textit{solc} compiler (v0.4.21) on an Intel Core i7 at 2.9 GHz and 16 Gb RAM, under Mac OS operating system. All applied statistical tests were run using IBM SPSS Statistics v.25 to decide the rejection or acceptance of formulated null hypotheses. To help selecting the proper statistical tests, we performed normal probability plots on the experimental data using p-p and q-q plots and residual analysis to gain insights into the distribution of the data. In view of the dependent variables and the results of normality plots, we utilized the ANOVA test which is solid and reliable. In all hypothesis testing, a 5\% significance level was chosen, henceforth we acknowledge a 5\% likelihood of committing a type-I-error that is rejecting a null hypothesis when it is sure. 

\section{experimental Design and Validity}
We designed our experiments as a randomized block with no human subjects included. Each smart contract was utilized specifically once to measure the impact of each treatment (tool). In other words, each of four security testing tools were randomly applied on the same ten smart contracts. This design helped to divide the variability into variability due to tools (i.e. treatments) and variability due to smart contracts (i.e. blocks). Thus, the impact of the tools could be analyzed without meddling from the impact of contracts covering the result of the investigations. The statistical analysis was based on the formal model \cite{parizi2011a} underlying ANOVA for a randomized block design of $y_{ij} = \lambda + a_i + b_j + \varepsilon_{ij}$, where $y_{ij}$ is the effectiveness or the accuracy score for contract \textit{i} given treatment \textit{j}, $\lambda$ is the overall mean effectiveness or accuracy score (i.e. $\mu_t$ or $\theta_t$), $a_i$ is the impact of contract \textit{i} (\textit{i}=1,...,10), and $b_j$ is the impact of treatment \textit{j} (\textit{j}=1,...,4). 

The threats to validity of this empirical research are as per the following. Concerning external validity, we used a subset of well-known open-source smart contracts from the blockchain community in assessing and testing smart contracts. The chosen smart contracts are generally suitable in terms of size and the number of functionalities. Without a doubt, they are sufficiently enormous to be reasonable and make the experiment practical. Moreover, the object programs contain diverse types of common vulnerabilities. As for the security testing tools, the chosen ones are the most commonly cited and used tools available in the literature at the time of this experiment. The threats to conclusion validity are concerned with the statistical analysis underling the conclusions. We selected our statistical tests based a thorough pre-analysis of the normality assessment of experiment's data (using p-p and q-q plots). As prescribed by plots, ANOVA and LSD (least significant difference) tests were chosen to investigate the null hypotheses ($H_{0Eff}$ and $H_{0Acc}$), which are solid. All the assumptions required by ANOVA test were further met. Therefore, the error rate would not be evident and the threat to conclusion validity could be minor.

\section{Results and Discussion}
The collected data are analyzed and discussed with respect to each research question in this section. In addition to descriptive analysis of data, we further perform hypothesis testing using ANOVA and LSD statistical tests to report on the significance of the results. 

\subsection{Analysis of Effectiveness (RQ1)}
The main visualization technique that we used to contrast the performance of tools was the Receiver Operating Characteristic (ROC\footnote{http://www.statisticshowto.com/c-statistic/}) analysis. The ROC curve shows how the recall vs. precision relationship changes, and we can learn a lot about a tool accuracy from this analysis. A ROC curve plots the true positive rate on the y-axis while the false positive rate is shown on the x-axis. The true positive rate (TPR) is the recall (sensitivity) and the false positive rate (FPR) is 1- specificity (i.e. $\frac{FP_c}{FP_c + TN_c}$ the probability of a false alarm by a tool on contract \textit{c}).

Following OWASP benchmark project\footnote{https://www.owasp.org/index.php/Benchmark}, Figure 3 shows the interpretation guide for ROC plots. Each plot shows a point plotted on the chart which provides a visual indication of how well a tool did in security vulnerability detection task.

\begin{figure}[h]
\centering
\includegraphics[scale=0.5]{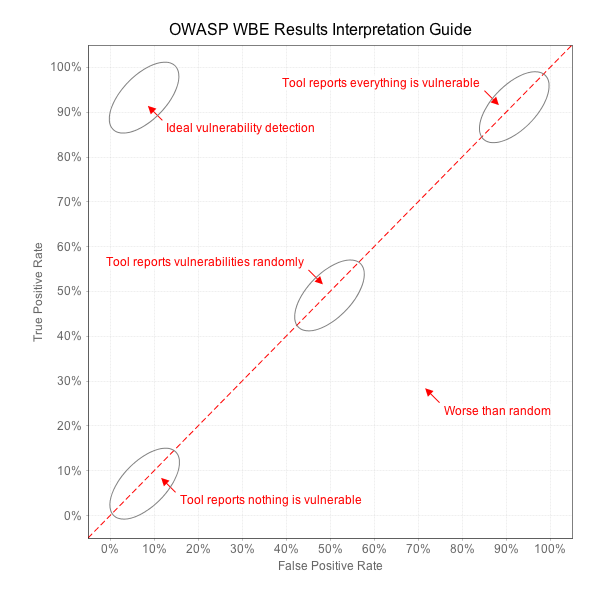}
\caption{Interpretation guide}
\end{figure}

As it can be seen from Figure 4, Oyente tool is not a good instrument since it is far from top left zone (ideal tool). This tool shows two breaking points when it crossed the guessing line and scored lower than random possibilities. This could be a strong disadvantage of this analyzer. Additionally, as we observed Oyente raised several outdated vulnerabilities that deal with class-stack depth as well as various falsely transaction-ordering dependence (TOD) alarms. 

\begin{figure}[h]
\centering
\includegraphics[scale=0.5]{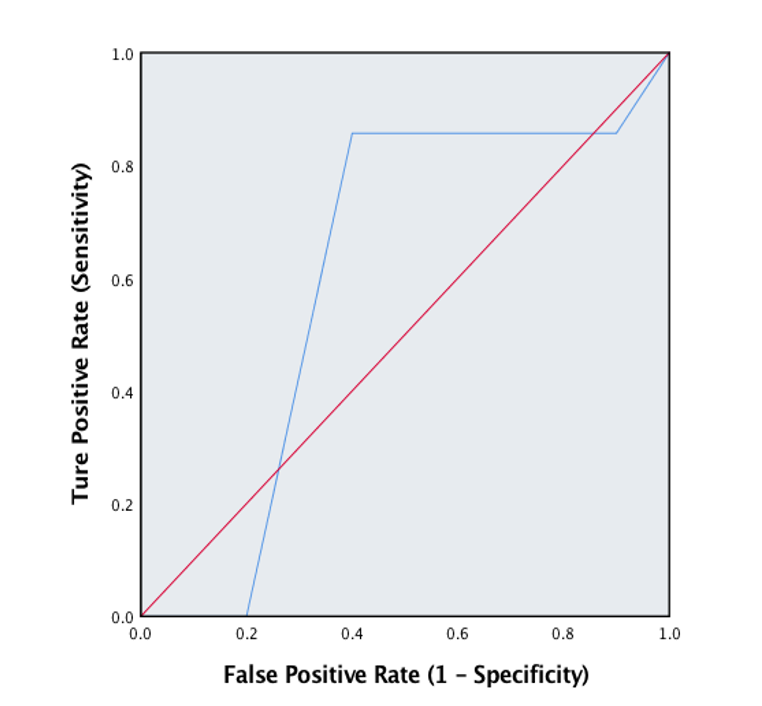}
\caption{ROC plot for Oyente tool}
\end{figure}

Figure 5 shows the ROC plot for the Securify tool. Securify showed a steady performance throughout the trials. It had a large number of false positives, but is still catching more threats that Oyente was missing.

\begin{figure}[h]
\centering
\includegraphics[scale=0.5]{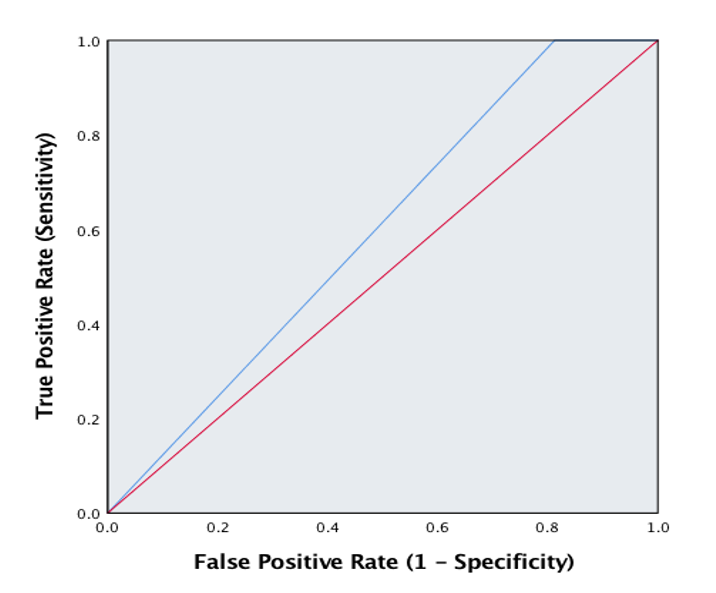}
\caption{ROC plot for Securify tool}
\end{figure}

Figure 6 shows the ROC plot for the Mythril tool. By looking at the graph, it is obvious that Mythril was able to detect more vulnerabilities compared to Oyente and Securify tools. It can be concluded that Mythril has a reasonable sensitivity and can be considered a trustworthy tool. 

\begin{figure}[h]
\centering
\includegraphics[scale=0.5]{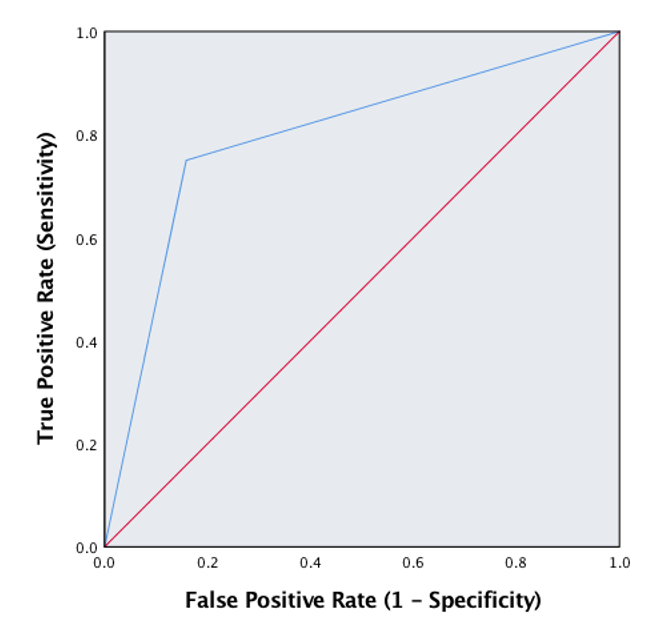}
\caption{ROC plot for Mythril tool}
\end{figure}

\begin{figure}[h]
\centering
\includegraphics[scale=0.5]{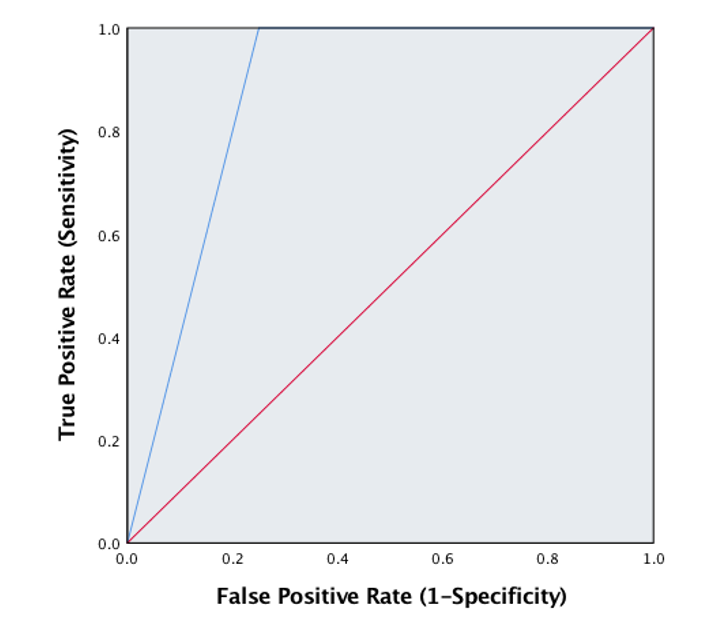}
\caption{ROC plot for SmartCheck tool}
\end{figure}

As it can be seen from Figure 7, SmartCheck showed a high degree of sensitivity compared to all other tested tools. Although, there was a steady ramp in the graph, this could be due to poor design in some of the contracts' logics.

Figure 8 contrast the percentages of effectiveness obtained from the testing tools over the ten smart contracts. The x-axis represents the ten smart contracts used and the y-axis signifies the effectiveness score for each single tool applied as result of the experiment. Neither tool catches everything, but it is clear that SmartCheck achieved much better score compared to its peers. Overall, the $Eff$ values of all smart contracts (10 out of 10) for SmartCheck tool have the highest figures.

\begin{figure}[h]
\includegraphics[scale=0.4]{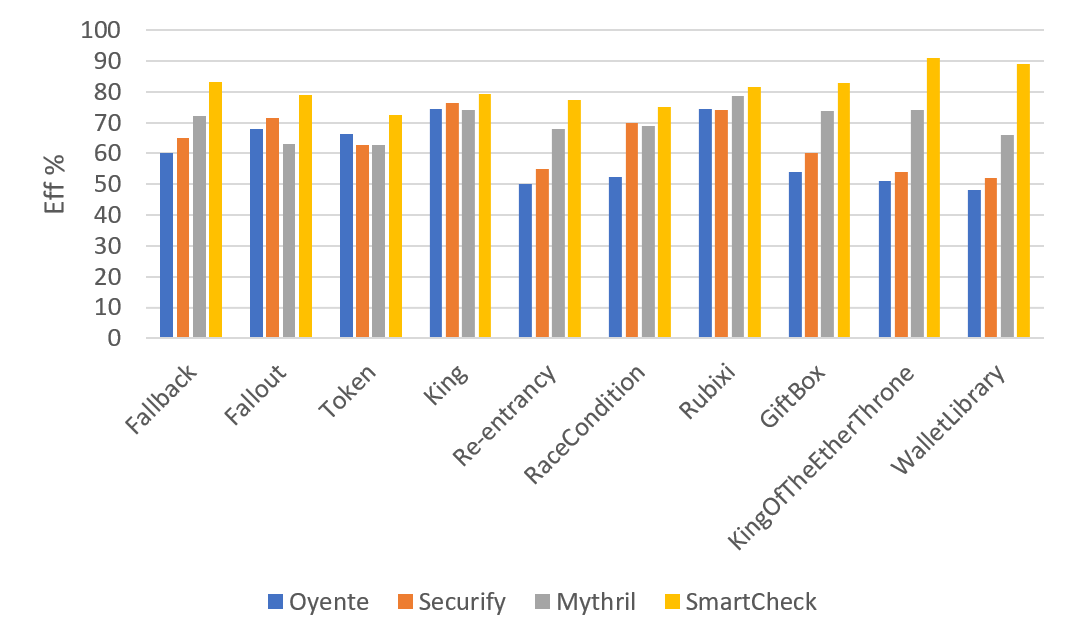}
\caption{Comparison of the vulnerability detectability of tools across all smart contracts}
\end{figure}

\begin{figure}[h]
\includegraphics[scale=0.4]{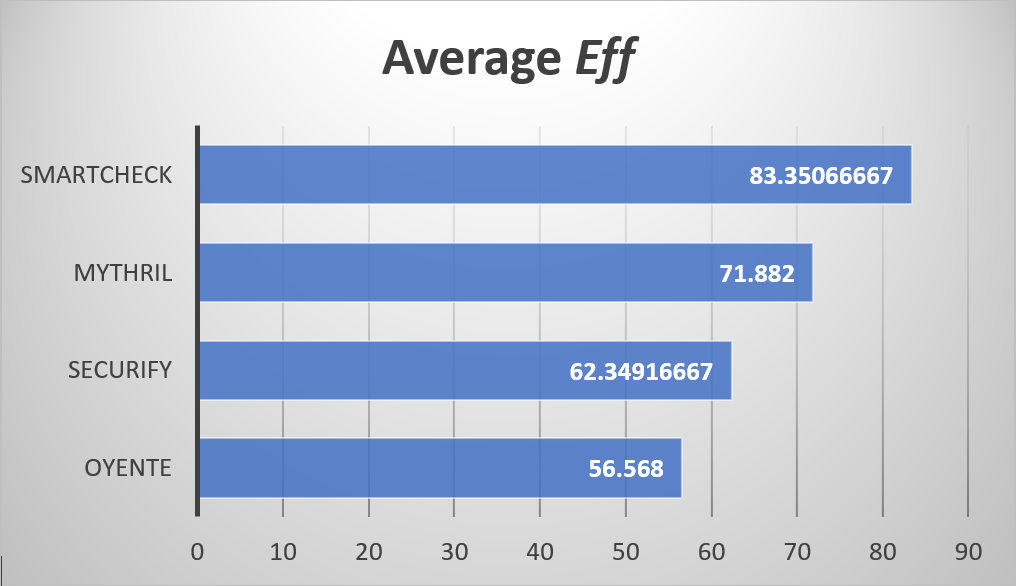}
\caption{Comparison of the average effectiveness by security testing tools}
\end{figure}

Figure 9 illustrates the total mean of the vulnerability detection effectiveness each tool from all ten smart contracts. The y-axis shows the tools and the x-axis shows the percentage of the average effectiveness. As it can be seen from the figure, SmartCheck accounted for the highest effectiveness in detection ability while the rest of tools had much lower statistics in all.

The provided descriptive comparisons (Figure 8 and Figure 9) intuitively give some insights into the effectiveness of the automated security testing tools in which not all data are identical (i.e.  $\mu_{Oyente} \neq \mu_{Securify} \neq \mu_{Mythril} \neq \mu_{SmartCheck}$). While SmartCheck demonstrated better results, this could make it conceivable to indicate that a distinction in vulnerability detection ability was noticed. Yet, it would not be adequate to reach conclusions, including rejecting the null hypothesis ($H_{0Eff}$), as it still requires conducting the statistical tests (given below) to decide the significant of the differences.

\textbf{Hypothesis testing ($H_{0Eff}$):} Table 3 shows the results of ANOVA test on the effectiveness of the tools. Given Table 3 and based on the decision rule, reject $H_0$ if p-value < $\alpha$ (whereas $0.0003 < 0.05$). Therefore, the null hypothesis ($H_{0Eff}$) is rejected and we can accept the alternative hypothesis, $H_{1Eff}$. In light of this analysis, the conclusion was drawn that there is a statistically significant distinction in the vulnerability capability of security testing tools for smart contracts. 

\begin{table}[htbp]
  \centering
  \caption{Results of ANOVA test on the effectiveness of the tools}
    \begin{tabular}{p{3.39em}ccccc}
    \toprule
    \multicolumn{1}{c}{} & \multicolumn{1}{p{3.835em}}{\textbf{Sum of Squares}} & \multicolumn{1}{p{1.11em}}{\textbf{df}} & \multicolumn{1}{p{3.335em}}{\textbf{Mean Square}} & \multicolumn{1}{p{2.335em}}{\textbf{F}} & \multicolumn{1}{p{2.835em}}{\textbf{Sig.}} \\
    \midrule
    Between Groups & 1491.176 & 3     & 497.059 & 8.214  & 0.0003 \\
    \midrule
    Within Groups & 1936.413 & 36    & 60.513 &       &  \\
    \midrule
    Total & 3427.590 & 39    &       &       &  \\
    \bottomrule
    \end{tabular}%
  \label{tab:addlabel}%
\end{table}%

\begin{table*}[htbp]
  \centering
  \caption{LSD multiple comparisons on the effectiveness of tools}
    \begin{tabular}{cp{5.5em}ccccc}
    \toprule
    \multicolumn{1}{c}{\multirow{2}[4]{*}{(I) Factor}} & \multirow{2}[4]{*}{(J) Factor} & \multirow{2}[4]{*}{Mean Difference (I-J)} & \multirow{2}[4]{*}{Std. Error} & \multirow{2}[4]{*}{Sig.} & \multicolumn{2}{p{15em}}{95\% Confidence Interval} \\
\cmidrule{6-7}          & \multicolumn{1}{c}{} &       &       &       & \multicolumn{1}{p{7.5em}}{Lower Bound} & \multicolumn{1}{p{7.5em}}{Upper Bound} \\
    \midrule
    \multicolumn{1}{c}{\multirow{3}[2]{*}{Oyente}} & Securify & -3.41111 & 3.66706 & 0.359 & -10.8807 & 4.0584 \\
          & Mythril & -6.88111 & 3.66706 & 0.07  & -14.3507 & 0.5884 \\
          & SmartCheck & -17.19111* & 3.66706 & 0.002 & -24.6607 & -9.7216 \\
    \midrule
    \multicolumn{1}{c}{\multirow{3}[2]{*}{Securify}} & Oyente & 3.41111 & 3.66706 & 0.359 & -4.0584 & 10.8807 \\
          & Mythril & -3.47 & 3.66706 & 0.351 & -10.9396 & 3.9996 \\
          & SmartCheck & -13.78* & 3.66706 & 0.001 & -21.2496 & -6.3104 \\
    \midrule
    \multicolumn{1}{c}{\multirow{3}[2]{*}{Mythril}} & Oyente & 6.88111 & 3.66706 & 0.07  & -0.5884 & 14.3507 \\
          & Securify & 3.47  & 3.66706 & 0.351 & -3.9996 & 10.9396 \\
          & SmartCheck & -10.31* & 3.66706 & 0.008 & -17.7796 & -2.8404 \\
    \midrule
    \multicolumn{1}{c}{\multirow{3}[2]{*}{SmartCheck}} & Oyente & 17.19111* & 3.66706 & 0.002 & 9.7216 & 24.6607 \\
          & Securify & 13.78* & 3.66706 & 0.001 & 6.3104 & 21.2496 \\
          & Mythril & 10.31* & 3.66706 & 0.008 & 2.8404 & 17.7796 \\
    \bottomrule
    \multicolumn{7}{p{39.665em}}{* The mean difference is significant at the 0.05 level.} \\
    \end{tabular}%
  \label{tab:addlabel}%
\end{table*}%

Furthermore, a post-hoc statistical test by means of least significant difference (LSD) was performed to help ascertain where the differences between tools lie and find the most effective security testing tool. Because, the ANOVA tests alone was not able to indicate which specific testing tools were significantly different from each other.

The results of this particular LSD test are shown in Table 4. For each pair of the tools the mean difference between their $Eff$ scores (Eq.1), the standard error of the difference, the significance level of the difference, and a 95\% confidence interval are shown. Significant outcomes are set apart with a mark (*) by SPSS software. In this case, SmartCheck tool demonstrated a significant difference compared to all other tools (Sig. $< 0.05$).

\subsection{Analysis of Accuracy (RQ2)}
Only achieving the highest effectiveness rate at any precision would not be a veritable pointer of best performance when it comes to real-world applications and tools. In this manner, it would be as imperative to investigate the accuracy scores obtained by the tools concerning false alarms to arrive at a more solid assessment. From a technical perspective, this two-facet evaluation would encourage fellow researchers and developers to pick the well-fitted tool by observing the trade-off between the effectiveness and the accuracy. The experiment results related to this aspect of analysis are presented in Figure 10 and Figure 11.

\begin{figure}[h]
\centering
\includegraphics[scale=0.4]{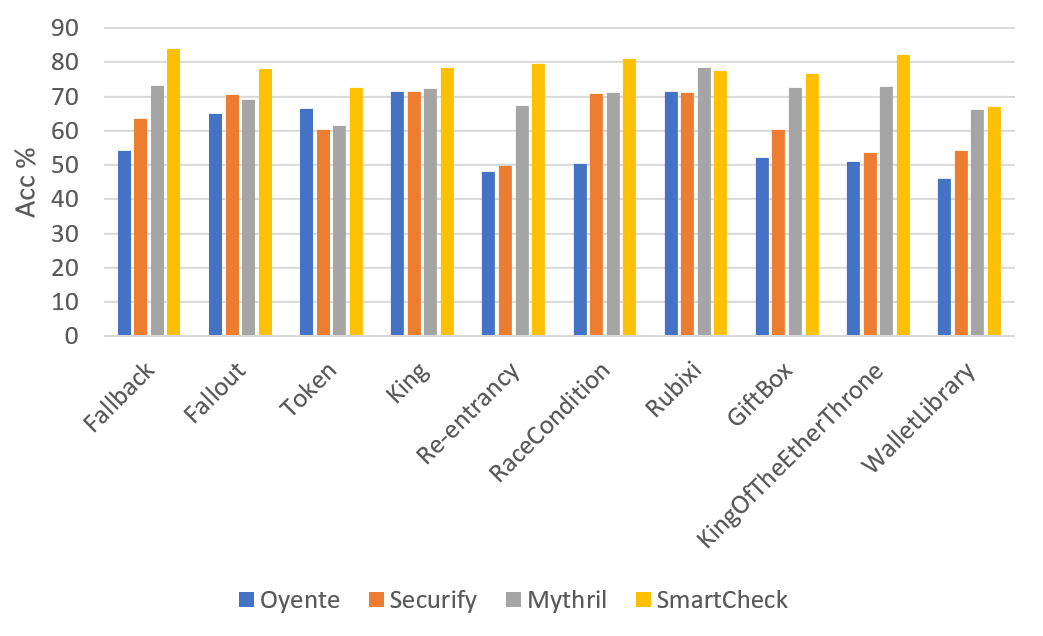}
\caption{Comparison of the accuracy of tools across all smart contracts}
\end{figure}

Figure 10 shows the percentages of accuracy scores obtained for the  tools over the ten smart contracts. The x-axis shows the ten smart contracts and the y-axis signifies the percentage of accuracy for each tool in the experiment. From the figure, Mythril and SmarkCheck tools almost outperformed all other peer tools in all ten smart contracts with recording highest accuracy scores.

\begin{figure}[h]
\centering
\includegraphics[scale=0.4]{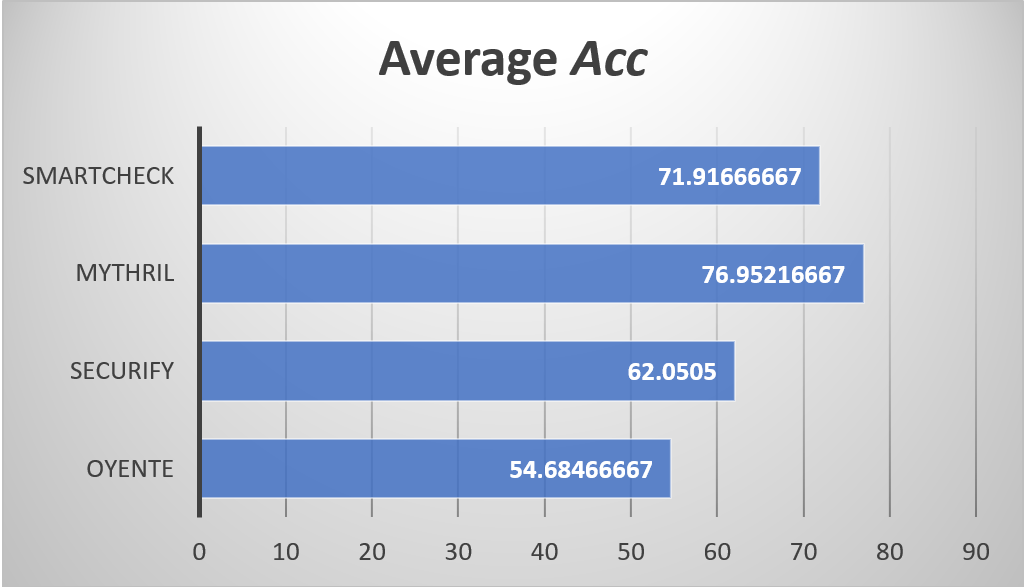}
\caption{Comparison of the average accuracy obtained by security testing tools}
\end{figure}

\begin{table}[htbp]
  \centering
  \caption{Results of ANOVA test on the accuracy of the tools}
    \begin{tabular}{p{3.39em}ccccc}
    \toprule
    \multicolumn{1}{c}{} & \multicolumn{1}{p{3.835em}}{\textbf{Sum of Squares}} & \multicolumn{1}{p{1.11em}}{\textbf{df}} & \multicolumn{1}{p{3.335em}}{\textbf{Mean Square}} & \multicolumn{1}{p{2.835em}}{\textbf{F}} & \multicolumn{1}{p{2.835em}}{\textbf{Sig.}} \\
    \midrule
    Between Groups & 2348.290 & 3     & 782.763 & 14.909 & 0.0002	 \\
    \midrule
    Within Groups & 1890.046 & 36    & 52.501 &       &  \\
    \midrule
    Total & 4238.336 & 39    &       &       &  \\
    \bottomrule
    \end{tabular}%
  \label{tab:addlabel}%
\end{table}%

\begin{table*}[htbp]
  \centering
  \caption{LSD multiple comparisons on the accuracy of the tools}
    \begin{tabular}{p{4.055em}cccccc}
    \toprule
    \multirow{2}[4]{*}{(I) Factor} & \multicolumn{1}{c}{\multirow{2}[4]{*}{(J) Factor}} & \multicolumn{1}{c}{\multirow{2}[4]{*}{Mean Difference (I-J)}} & \multicolumn{1}{c}{\multirow{2}[4]{*}{Std. Error}} & \multicolumn{1}{c}{\multirow{2}[4]{*}{Sig.}} & \multicolumn{2}{p{15em}}{95\% Confidence Interval} \\
\cmidrule{6-7}    \multicolumn{1}{c}{} &       &       &       &       & \multicolumn{1}{p{7.5em}}{Lower Bound} & \multicolumn{1}{p{7.5em}}{Upper Bound} \\
    \midrule
    \multicolumn{1}{c}{\multirow{3}[2]{*}{1.00}} & 2.00    & \multicolumn{1}{p{6.5em}} {-5.01500} & 3.24041 & 0.13  & -11.5869 & 1.5569 \\
    \multicolumn{1}{c}{} & 3.00     & \multicolumn{1}{p{6.5em}}{-20.12500*} & 3.24041 & 0     & -26.6969 & -13.5531 \\
    \multicolumn{1}{c}{} & 4.00     & \multicolumn{1}{p{6.5em}}{-12.90200*} & 3.24041 & 0     & -19.4739 & -6.3301 \\
    \midrule
    \multicolumn{1}{c}{\multirow{3}[2]{*}{2.00}} & 1.00     & \multicolumn{1}{p{6.5em}} {5.01500} & 3.24041 & 0.13  & -1.5569 & 11.5869 \\
    \multicolumn{1}{c}{} & 3.00     & \multicolumn{1}{p{6.5em}}{-15.11000*} & 3.24041 & 0     & -21.6819 & -8.5381 \\
    \multicolumn{1}{c}{} & 4.00     & \multicolumn{1}{p{6.5em}}{-7.88700*} & 3.24041 & 0.02  & -14.4589 & -1.3151 \\
    \midrule
    \multicolumn{1}{c}{\multirow{3}[2]{*}{3.00}} & 1.00     & \multicolumn{1}{p{6.5em}}{20.12500*} & 3.24041 & 0     & 13.5531 & 26.6969 \\
    \multicolumn{1}{c}{} & 2.00     & \multicolumn{1}{p{6.5em}}{15.11000*} & 3.24041 & 0     & 8.5381 & 21.6819 \\
    \multicolumn{1}{c}{} & 4.00     & \multicolumn{1}{p{6.5em}}{7.22300*} & 3.24041 & 0.032 & 0.6511 & 13.7949 \\
    \midrule
    \multicolumn{1}{c}{\multirow{3}[2]{*}{4.00}} & 1.00     & \multicolumn{1}{p{6.5em}}{12.90200*} & 3.24041 & 0     & 6.3301 & 19.4739 \\
    \multicolumn{1}{c}{} & 2.00     & \multicolumn{1}{p{6.5em}}{7.88700*} & 3.24041 & 0.02  & 1.3151 & 14.4589 \\
    \multicolumn{1}{c}{} & 3.00     & \multicolumn{1}{p{6.5em}}{-7.22300*} & 3.24041 & 0.032 & -13.7949 & -0.6511 \\
    \midrule
    \multicolumn{7}{p{39.665em}}{* The mean difference is significant at the 0.05 level.} \\
    \end{tabular}%
  \label{tab:addlabel}%
\end{table*}%

On the average basis and interestingly, Mythril tool showed the highest accuracy score, though it had less effectiveness than SmartCheck. Overall, it is safe to say that there were differences observed in the accuracy of all the testing tools, but descriptive results could indicate that the Mythril and SmartCheck tools are more accurate than Oyente and Securify with less false alarms. The ANOVA and LSD statistical tests performed below show the significant of the differences. 

\textbf{Hypothesis testing ($H_{0Acc}$):} To factually exhibit the accuracy's differences between the tools under evaluation, the ANOVA was once again carried out to check the respected hypothesis ($H_{0Acc}$). Likewise with the past ANOVA test on the effectiveness, the confidence level of significance for the hypothesis testing was set to $\alpha = 0.05$ as well. Table 5 and Table 6 show the results of ANOVA statistical test and LSD respectively.

As it can be seen from Table 5, p-value ($0.0002$) is less than $\alpha = (0.05)$, thus the results suggest rejecting the null hypothesis $H_{0Acc}$ in favor of the alternative $H_{1Acc}$ at the $0.05$ significance level. Following the recommendation, the choice was made to dismiss the null hypothesis and as needs be, acknowledge the alternative hypothesis.

As we were keen to find out where the contrasts between the testing tools lie, a LSD test was additionally performed to indicate the differences in a pair-wise manner. For each couple of the tools the mean difference between their accuracy scores (Eq. (\ref{two})), the standard error of the difference, the significance level of the difference, and a 95\% confidence level are shown. Notably, Mythril and SmartCheck tools demonstrated significant differences in all pairs with the remaining tools (Sig. $< 0.05$). 

Finally, it can be concluded that the results of these experiments seemingly set up a trade-off between the vulnerability capability detection and the accuracy of smart contract-specific security testing tools in the Ethereum blockchain, likewise with classic trade-off between effectiveness and efficiency reported in the literature in all realms of testing.

\section{Related Work}
The nature of work given in this paper is generally related to security testing and vulnerability assessment of smart contracts on the blockchain. This section presents the related work (including approaches and tools) and secondary studies (including experiments and surveys) in this area. 

Most recently, Parizi et al. \cite{parizi2018b}, conducted an empirical analysis of smart contract programming languages based on usability and security from new developers point of view. They selected three programming languages for their experiment, i.e. Solidity, Pact\footnote{http://kadena.io/docs/Kadena-PactWhitepaper.pdf} (a high-level language for the Kadena\footnote{http://kadena.io/} platform) and Liquidity\footnote{https://github.com/OCamlPro/liquidity/blob/master/docs/liquidity.md} (a high-level language for the Tezos\footnote{https://tezos.com/} platform). The results of their experiment indicated that new contract developers found Solidity to be the most efficient language with the highest usability score and the shortest average implementation times. But, it was found that 73.33\% of implemented Solidity contracts had security vulnerabilities, while no known security vulnerabilities were found in contracts implemented with Pact and Liquidity. In conclusion, the study concluded that although Solidity is the most usable language to a new developer, it is also the most vulnerable to malicious attacks as new developers tend to leave behind security vulnerabilities which can leave the contracts insecure.

Destefanis et al. \cite{destefanis2018a} advocated the need for a discipline of Blockchain Software Engineering (BOSE), addressing the security issues posed by smart contract programming and other applications running on the Ethereum blockchain. The authors presented a case study of the Parity\footnote{https://www.parity.io/} wallet's smart contract library, where poor programming practices led to a situation where an anonymous user was able to freeze about 500K Ether (150M USD) in November 2017. The analysis of the case led to the authors concluding that vulnerability of the library was mainly due to a negligent programming activity rather than a problem in the Solidity language.

Atzei et al. \cite{atzei-a} provided a systematic exposition of the security vulnerabilities of Ethereum and Solidity. The authors presented a taxonomy of causes of vulnerabilities, classifying them on three levels namely Solidity (Call to the unknown, Gasless send, Exception disorders, Type casts, Reentrancy, keeping secrets), EVM (Immutable bugs, Ether lost in transfer, Stack size limit) and Blockchain (Unpredictable state, Generating randomness, Time constraints). Additionally, the authors accompanied their taxonomy with actual attacks which exploit detected vulnerabilities except Type casts, Ether lost in transfer and generating randomness vulnerabilities. To conclude their survey, the authors recommended the process of formal verification of smart contracts to ensure that the intended behavior and the actual behavior of the smart contracts are exactly the same. Moreover, the authors suggested the use of Turing-incomplete, human-readable languages for formal verification as the choice of Turing complete languages limits formal verification.

While the above-mentioned works primarily focused on comparing and highlighting the possible security vulnerabilities with Ethereum and Solidity, the works mentioned below focus on minimizing and mitigating security vulnerabilities with various approaches for verification of smart contracts \cite{magazzeni2017a} on the Ethereum blockchain. 

Abdellatif and Brousmich \cite{abdellatif2018a} proposed a semantics-based formal method approach for verfication of smart contracts and blockchain properties. The authors modeled a smart contract's behavior and interactions with its hosting environment by applying this approach through an illustrative example. They simulated these behaviors in the BIP (Behavior Interaction Priorities) framework \cite{basu2006a}, equipped with a series of runtime verification and simulation engines \cite{bensalem2009a}. In this study, results were analyzed using a SMC (Statistical Model Checking) \cite{legay2010a} tool, which allowed the authors to reveal scenarios where the smart contract behaviors could be compromised by malicious users.

Bhargavan et al. \cite{bhargavan2016a} designed a framework to examine and verify the runtime safety and the functional correctness of the Ethereum contracts by translation to F*, a functional programming language aimed at program verification. The authors proposed two prototype tools which translate the contracts to F* programs for developing more secure end-contracts. In a similar work, Grishchenko et al. \cite{grishchenko2018a} presented the first complete small-step semantics of EVM bytecode, which they formalized in the F* proof assistant based on a combination of hyper- and safety properties, obtaining executable code that they successfully validated against the official Ethereum test suite. 

Mavridou and Laszka \cite{mavridou2017a} introduced \textit{FSolidM}\footnote{https://github.com/anmavrid/smart-contracts} , a framework rooted in rigorous semantics for designing smart contracts as Finite State Machines (FSM). The authors presented a tool for creating FSM on a highly usable GUI and for automatically generating Ethereum smart contracts. The authors \cite{mavridou2018a} provided a demonstration of the \textit{FSolidM} tool in a later work. 

Kalra et al. \cite{kalra2018a} proposed a framework 'ZEUS' to verify the correctness and validate the fairness of smart contracts. ZEUS leverages both abstract interpretation and symbolic model checking, along with the power of constrained horn clauses to quickly verify contracts for safety. The authors built a prototype of ZEUS for Ethereum and Fabric\footnote{https://www.hyperledger.org/projects/fabric} blockchain platforms and evaluated it with smart contracts. The evaluation found 94.6\% of contracts (containing cryptocurrency worth more than USD 0.5 billion) to be securely vulnerable. 

Breindenbach et al. \cite{breindenbach2018a} proposed the \textit{Hydra} Framework, which is a principled approach for modeling and administering bug bounties that incentivize bug disclosure. The framework transforms programs via N-of-N-version programming (NNVP), a variant of classical N-version programming \cite{avizienis1995a} that runs multiple independent program instances. The \textit{Hydra} Framework was applied to sample smart contracts and it was concluded it could greatly amplify the power of bounties to incentivize bug disclosure by economically rational adversaries. 

Lastly, Idelberger et al. \cite{idelberger2016a} investigated the possibilities of using logic-based smart contracts on blockchain systems to boost their general safety and security. The authors demonstrated that a logical-based approach could beneficially complement its procedural counterpart with respect to the negotiation, formation, storage/notarizing, enforcement, monitoring and activities related to dispute resolution. It was proven that the logic and procedural approaches are not incompatible, contrarily, they have the potential to advantageously complement each other for more quality contracts.

In summary, it can be said that sound empirical studies in the field of smart contract security is currently lagging as compared to primary studies. To the extent of our knowledge, the work presented in this paper was the first of its kind that statistically analyzes and evaluates security analysis tools for smart contracts based on the effectiveness in vulnerability detection and the accuracy.

\section{Conclusions}
In this paper, we presented a comprehensive empirical evaluation of open source automatic security analysis tools for the security vulnerability detection of Ethereum smart contracts written in Solidity. We tested those tools on ten real-world smart contracts from both vulnerability effectiveness and accuracy of true detection viewpoints.  The results of our experiments showed that SmartCheck tool is statistically more effective than the other automated security testing tools at 95\% significance level (p $< 0.05$). Concerning the accuracy, Mythril was found to be significantly (p $< 0.05$) accurate with issuing the lowest number of false alarms among peer tools. These results could imply that SmartCheck could currently be the most effective static security testing tool for Solidity smart contracts on the Ethereum blockchain but perhaps less accurate than Mythril. 

As a general conclusion, our work indicates that research on the empirical knowledge evaluation of security testing for smart contracts is scarce in the literature, especially in relation to IoT, as perhaps this domain is still in a state of infancy. Hence, our work contributed towards filling this gap by providing: (1) a fine-grained methodology to conduct such empirical study for future use by fellow researchers, (2) comparable experimental results on the state-of-the-art smart contracts security testing tools, and (3) statistical tests and constructive insights into the challenges associated with testing smart contracts. We hope our work motivates researchers and developers to come up with more new and innovative ideas, frameworks, and tools that would result in writing safer, secure and vulnerability-immune smart contracts in the future. In our view, IoT combined with smart contracts on blockchains are helpful in building more trustworthy and secure networks, and it would appear to hold great promise for future IoT security development.

\bibliographystyle{ACM-Reference-Format}
\bibliography{references}

\end{document}